\documentclass[letterpaper,twocolumn]{jpsj3}
\usepackage{txfonts}

\usepackage{graphicx}
\usepackage{dcolumn}
\usepackage{bm}
\usepackage{physics}
\usepackage{qcircuit}
\usepackage{here}

\newcommand{\arrep}[1]{\ar@{-}@/^/[#1]|-{\mbox{ $L$ times }}}

\title{Fourier Analysis of Parameterized Quantum Circuits and the Barren Plateau Problem}

\author{Shun Okumura$^{1}$ and Masayuki Ohzeki$^{1,2,3,4}$}
\inst{
$^{1}$Graduate School of Information Sciences, Tohoku University, Miyagi 980-8564, Japan \\
$^{2}$Department of Physics, Science Tokyo, Meguro, Tokyo 152-8551, Japan\\
$^{3}$Research and Education Institute for Semiconductors and Informatics, Kumamoto University, Chuo, Kumamoto 860-0862, Japan\\
$^{4}$Sigma-i Co., Ltd., Tokyo, 108-0075, Japan
}

\abst{
We reveal a universal link between the Fourier representation of parameterized quantum circuits (PQCs) and the barren plateau (BP) problem. 
Because PQCs depend periodically on circuit parameters, their outputs can be analyzed via Fourier expansion. 
Using Parseval’s identity, we show that the total spectral power—the sum of squared Fourier coefficients—is exponentially suppressed with qubit number or circuit depth under BP conditions, offering a spectral interpretation of gradient vanishing. 
This framework requires only parameter periodicity, independent of initialization or ansatz assumptions, and thus generalizes the understanding of BPs to all periodic PQCs. 
Numerical results with hardware-efficient circuits support this correspondence, establishing a unified perspective that connects expressibility, spectral structure, and trainability in quantum machine learning.
}

\begin{document}
\maketitle

\textit{Introduction.}
Quantum computers promise exponential or polynomial speedups over classical counterparts, exemplified by Shor’s factoring algorithm \cite{1}, the quantum linear-systems algorithm \cite{2}, and quantum chemical simulations \cite{3}. 
However, these approaches require fault-tolerant quantum hardware, still beyond current capabilities. 
Current research instead targets noisy intermediate-scale quantum (NISQ) devices \cite{4}, which combine limited qubit counts with hybrid quantum–classical optimization. 
Parameterized quantum circuits (PQCs) serve as their core, forming the basis of the variational quantum eigensolver (VQE) \cite{5}, quantum approximate optimization algorithm (QAOA) \cite{6}, and quantum neural networks (QNNs) \cite{7,8}, applied across fields such as finance \cite{9_1,9_2} and quantum chemistry \cite{9_3,9_4}.

A major obstacle to training PQCs is the \textit{barren plateau} (BP) problem, where the cost-function gradient variance decays exponentially with qubit number \cite{10,11}, producing flat optimization landscapes and rendering gradient-based methods ineffective. 
The BP can occur regardless of the optimization strategy, including higher-order and gradient-free methods \cite{12,13}—and although mitigation strategies exist, a unified theoretical explanation remains lacking.

We employ Fourier analysis to provide new insight into this problem. 
Since PQCs depend periodically on circuit parameters, their outputs admit Fourier expansion. 
This periodicity arises generally from unitaries of the form $V(\theta)=e^{-i\theta H}$ for any Hermitian $H$, not only Pauli rotations, making the approach broadly applicable. 
While Fourier techniques have been used to study expressibility, generalization, and learning capacity in QNNs \cite{15,16,17_1,17_2}, their direct link to the barren plateau phenomenon has not been established.

Here, we reveal a fundamental spectral connection between the Fourier representation of PQCs and barren plateaus. 
Using Parseval’s identity, we show that (i) the sum of squared Fourier coefficients—interpreted as total spectral power—is exponentially suppressed under BP conditions, and (ii) this suppression cannot be avoided merely by altering parameter initialization. 
Because Parseval’s identity depends only on periodicity, the framework applies universally to periodic PQCs. 
Together with numerical verification, this establishes a spectral interpretation of the BP problem, offering a general theoretical perspective on trainability in quantum machine learning.

\textit{Parameterized Quantum Circuits.}
A single-parameter unitary gate is defined as
\begin{equation}
V(\theta) = e^{-i\theta H} = I\cos\theta - iH\sin\theta,
\end{equation}
where $H$ is Hermitian and $\theta$ is real. 
Its trigonometric form ensures $2\pi$-periodicity, 
\begin{equation}
V(\theta) = V(\theta + 2\pi),
\end{equation}
establishing the foundation for Fourier analysis. 
When $H^2 = I$, as for Pauli rotations, the period is $2\pi$; for a general $H = \sum_j \lambda_j |j\rangle!\langle j|$, periodicity holds with period $2\pi/|\lambda_{\max}-\lambda_{\min}|$. 
Since our analysis relies only on periodicity, all results apply to general parameterized unitaries up to rescaling of the period.

A general PQC (ansatz) is constructed as
\begin{equation}
U(\bm{\theta}) = \prod_{i=1}^{D} V_i(\theta_i) W_i,
\end{equation}
where $W_i$ are fixed unitaries such as Hadamard or CNOT gates. 
Depending on the task, PQCs yield two types of outputs: \textit{expectation-value} and \textit{probability}.

For the expectation-value form,
\begin{equation}
f(\bm{\theta}) = \Tr[U(\bm{\theta})^{\dagger}\rho U(\bm{\theta}) O], \label{f}
\end{equation}
with $\rho=\dyad{0}^{\otimes n}$ and observable $O$, $f(\bm{\theta})$ is bounded by $|f|\le1$ and $2\pi$-periodic in each $\theta_i$. 
In QNNs, data are encoded via an additional unitary $E(\bm{x})$:
\begin{equation}
f(\bm{x},\bm{\theta}) = \Tr[U(\bm{\theta})^{\dagger}E(\bm{x})^{\dagger}\rho E(\bm{x})U(\bm{\theta})O],
\end{equation}
where $E(\bm{x})$ acts as a feature map, influencing both expressibility and susceptibility to barren plateaus.

For probability-type outputs,
\begin{equation}
q_{\bm{b}}(\bm{\theta}) = \Tr[U(\bm{\theta})^{\dagger}\rho U(\bm{\theta})P_{\bm{b}}], \label{q}
\end{equation}
with $P_{\bm{b}}=\dyad{\bm{b}}$, as used in quantum circuit Born machines \cite{18} and kernel methods \cite{21}.

Gradients are efficiently computed by the parameter-shift rule \cite{7,20}:
\begin{equation}
\pdv{f(\bm{\theta})}{\theta_i}
=\frac{1}{2}\!\left[f\!\left(\bm{\theta}+\frac{\pi}{2}\bm{e}_i\right)
-f\!\left(\bm{\theta}-\frac{\pi}{2}\bm{e}_i\right)\!\right],
\end{equation}
avoiding the numerical instability of finite differences.  
Thus, the intrinsic periodicity and differentiability of PQCs make them naturally suited to Fourier analysis, forming the basis for the barren plateau study developed below.

\textit{Expressivity.}
Selecting an appropriate ansatz $U(\bm{\theta})$ is central to variational quantum algorithms \cite{27}. 
Expressivity quantifies how broadly a circuit explores the Hilbert space—specifically, how closely its ensemble of unitaries approximates the uniform (Haar) distribution \cite{22}. 

Formally, for $\mathbb{U} := \{U(\bm{\theta}) \mid \bm{\theta}\!\sim\! P\}$, the expressivity is defined by the deviation of its $t$-th moment operator from that of the Haar measure:
\begin{multline}
\epsilon_{\mathbb{U}}^{(t)}(\rho) :=
\Bigl\|
\!\int_{\mathrm{Haar}}\! d\mu(V)\, V^{\dagger \otimes t}\rho^{\otimes t}V^{\otimes t}
-\!
\int_{\mathbb{U}}\! dU\, U^{\dagger \otimes t}\rho^{\otimes t}U^{\otimes t}
\Bigr\|_1,
\end{multline}
where $\|\cdot\|_1$ is the trace norm. 
Smaller $\epsilon_{\mathbb{U}}^{(t)}(\rho)$ indicates higher expressivity; when it vanishes, the circuit forms a \textit{unitary $t$-design} \cite{28}, reproducing Haar averages for all degree-$t$ polynomials. 
Thus, a maximally expressive ansatz can, in principle, cover the full Hilbert space.

\textit{Barren Plateau Problem.}
High expressivity, while enabling broad state representation, can hinder trainability. 
When an ansatz $U(\bm{\theta})$ approaches a unitary 2-design, the gradient variance of the cost function decays exponentially with qubit number \cite{10,11}:
\begin{equation}
\mathrm{Var}_{\bm{\theta}}\!\left[\pdv{f(\bm{\theta})}{\theta_i}\right]=\mathcal{O}(b^{-n}),
\end{equation}
where $b>0$ and $n$ is the number of qubits. 
This exponential decay flattens the optimization landscape, rendering gradient-based training infeasible. 
Mitigation attempts include local measurements and nonuniform initialization \cite{23,24}, with similar concentration effects reported in quantum kernel methods \cite{25}. 

In our framework, the 2-design condition mainly concerns the ansatz $U(\bm{\theta})$, yet a data encoder $E(\bm{x})$ can cause equivalent gradient collapse if it approximates a 2-design across inputs. 
In either case, the output concentrates exponentially around a constant, erasing useful gradients. 
The following analysis connects this expressivity–trainability trade-off to the decay of Fourier coefficients, providing a spectral interpretation of the barren plateau phenomenon.

\textit{Main Results.}
Because the functions in Eqs.~(\ref{f}) and (\ref{q}) are periodic in each parameter, they can be expanded into Fourier series:
\begin{equation}
f(\bm{\theta}) = \sum_{|\bm{k}| \le \infty} c_{\bm{k}} e^{-i\bm{k}\cdot\bm{\theta}},
\end{equation}
where $c_{\bm{k}}$ are Fourier coefficients. Parseval’s identity relates their squared sum to the $L^2$ norm of $f$:
\begin{equation}
\sum_{|\bm{k}| \le \infty} |c_{\bm{k}}|^2 = \frac{1}{(2\pi)^d}\!\int_{[0,2\pi]^d}\!|f(\bm{\theta})|^2 d\bm{\theta}, \label{par}
\end{equation}
with $d$ the number of parameters.

For a composite circuit of $L$ parameterized blocks,
\begin{equation}
f(\bm{\theta}_1,\!\ldots\!,\bm{\theta}_L)
= \Tr\!\left[\!\prod_{l=1}^L U_l(\bm{\theta}_l)^{\dagger}\rho \!\prod_{l=1}^L\! U_l(\bm{\theta}_l)O\!\right]\!, \label{multi_f}
\end{equation}
we isolate the $i$th block by defining
\begin{align}
\rho(\bm{\theta}_{<i}) &= \prod_{l<i}\! U_l^\dagger(\bm{\theta}_l)\rho U_l(\bm{\theta}_l),\\
O(\bm{\theta}_{>i}) &= \prod_{l>i}\! U_l(\bm{\theta}_l)O U_l^\dagger(\bm{\theta}_l),
\end{align}
and treat $f$ as periodic in $\bm{\theta}_i$:
\begin{equation}
f(\bm{\theta})=\!\sum_{|\bm{k}| \le \infty}\! c_{\bm{k}}(\bm{\theta}_{\neg i}) e^{-i\bm{k}\cdot\bm{\theta}_i}.
\end{equation}
Applying Parseval’s identity gives
\begin{equation}
\sum_{|\bm{k}| \le \infty}\! |c_{\bm{k}}(\bm{\theta}_{\neg i})|^2
=\frac{1}{(2\pi)^{d_i}}\!\int_{[0,2\pi]^{d_i}}\!|f(\bm{\theta})|^2d\bm{\theta}_i. \label{par2}
\end{equation}

Interpreting the right-hand side as the second moment over a uniform distribution links it to the expressivity $\epsilon_{\mathbb{U}}^{(2)}(\rho)$:
\begin{multline}
\Bigg|\sum_{|\bm{k}| \le \infty}\!|c_{\bm{k}}|^2-\frac{1}{2^n+1}\Bigg|
\le \epsilon_{\mathbb{U}_i}^{(2)}\!\bigl(\rho(\bm{\theta}_{<i})\bigr).
\end{multline}
If $U_i(\bm{\theta}_i)$ forms a unitary 2-design, $\epsilon_{\mathbb{U}_i}^{(2)}=0$, yielding
\begin{equation}
\sum_{|\bm{k}| \le \infty}\! |c_{\bm{k}}|^2=\frac{1}{2^n+1}. \label{fm}
\end{equation}
Thus, the total spectral power decreases exponentially with qubit number $n$, and each $|c_{\bm{k}}|$ becomes exponentially small, flattening the landscape:
\begin{equation}
f(\bm{\theta}_1,\ldots,\bm{\theta}_L)\approx0. \label{f0}
\end{equation}
Expectation values concentrate near constants, and gradients vanish—an explicit spectral form of the barren plateau.

For probability-type outputs, an analogous expression holds:
\begin{equation}
\sum_{|\bm{k}| \le \infty}\!|c_{\bm{k}}|^2=\frac{1}{2^{n-1}(2^n+1)}.
\end{equation}
In both cases, increasing $n$ exponentially suppresses accessible spectral power, forcing the output distribution to concentrate and eliminating informative gradients.

These results show that Parseval’s identity bridges expressivity and trainability: barren plateaus arise from intrinsic spectral suppression, not from initialization choices, explaining why changing the parameter distribution cannot generally avoid them.

\textit{Numerical experiments.}
We numerically validate the spectral predictions using a minimal setup. 
A one-variable QNN is constructed where a hardware-efficient embedding (HEE) of depth $L$ encodes a scalar input $x$, followed by a shallow trainable layer $U(\bm{\theta})$ composed of single-qubit rotations. 
Both the HEE and the related hardware-efficient ansatz (HEA) are known to exhibit barren plateaus with increasing depth, and in this configuration the 2-design-like behavior arises mainly from the encoder $E^L(x)$ rather than $U(\bm{\theta})$.

We consider $n\in\{2,4,6,8\}$ qubits and depths $L\in\{5,10,15,20,25,30,35,40,45,50\}$. 
The observable is the global Pauli operator $Z^{\otimes n}$, and all qubits are measured in the $Z$ basis. 
Parameters are initialized as $\theta_j\sim\mathrm{Unif}[0,2\pi)$ and inputs as $x\sim\mathrm{Unif}[0,2\pi)$. 
Simulations were performed using \textsc{PennyLane} \cite{penny}. 
Figure~\ref{c1} shows the $L$-layer HEE and Fig.~\ref{c2} the full QNN, whose expectation is
\[
f(x,\bm{\theta})=\Tr\!\left[U(\bm{\theta})^\dagger E^L(x)^\dagger \rho\,E^L(x)U(\bm{\theta})\, Z^{\otimes n}\right].
\]

Because $f$ is $2\pi$-periodic in each parameter, we estimate the total spectral power (sum of squared Fourier coefficients) along a single parameter using a discrete Fourier transform on a uniform $[0,2\pi)$ grid. 
For each $(n,L)$ pair, $300$ random seeds (re-sampling $\bm{\theta}$ and $x$) are drawn, and the mean and variance of the spectral power are evaluated. 
This Monte Carlo estimate corresponds to the right-hand side of Eq.~(\ref{par2}) and should approach the 2-design value when $E^L(x)$ becomes sufficiently expressive.

\begin{figure}
\begin{center}
\includegraphics[width=70mm]{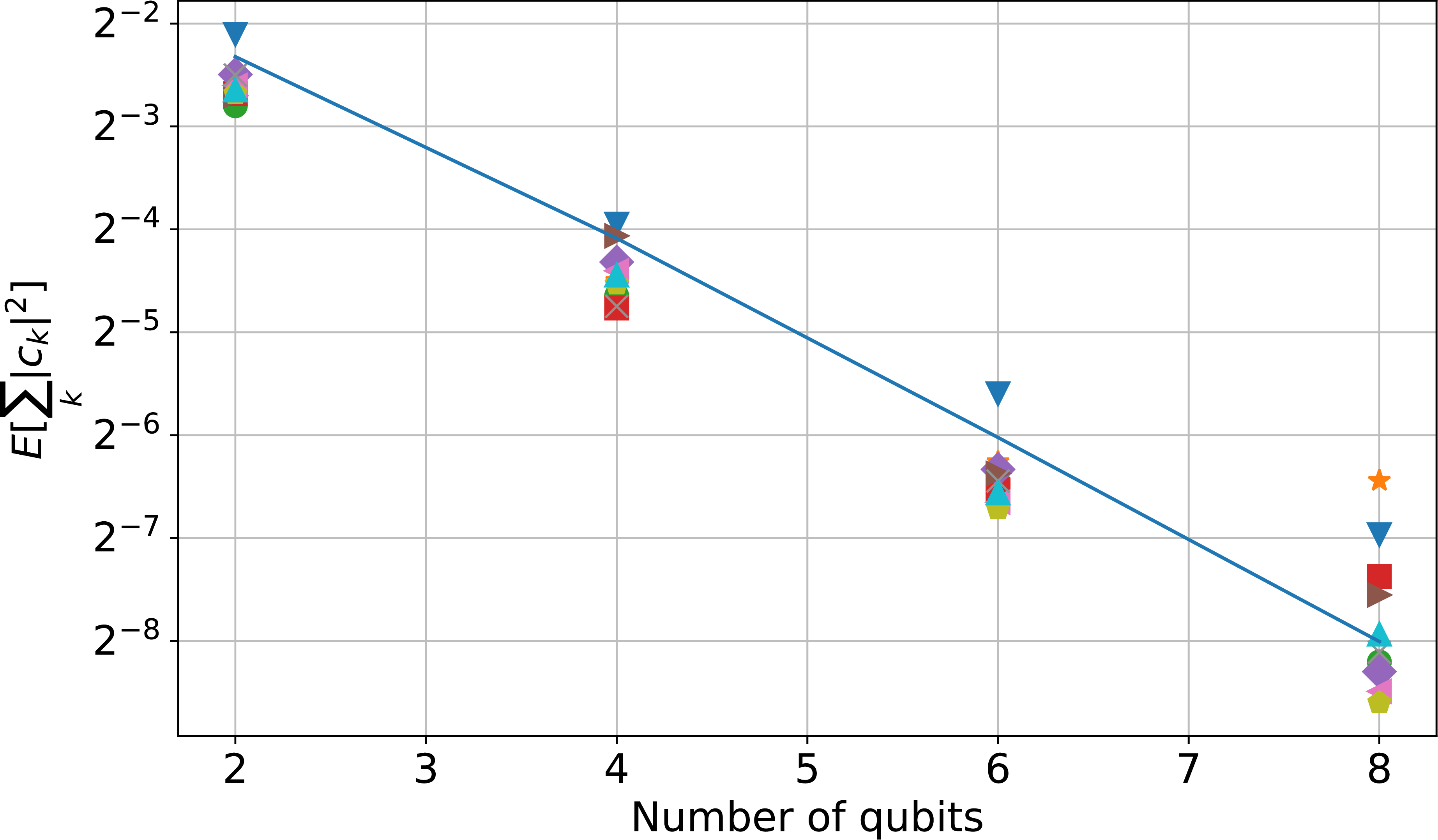}
\includegraphics[width=70mm]{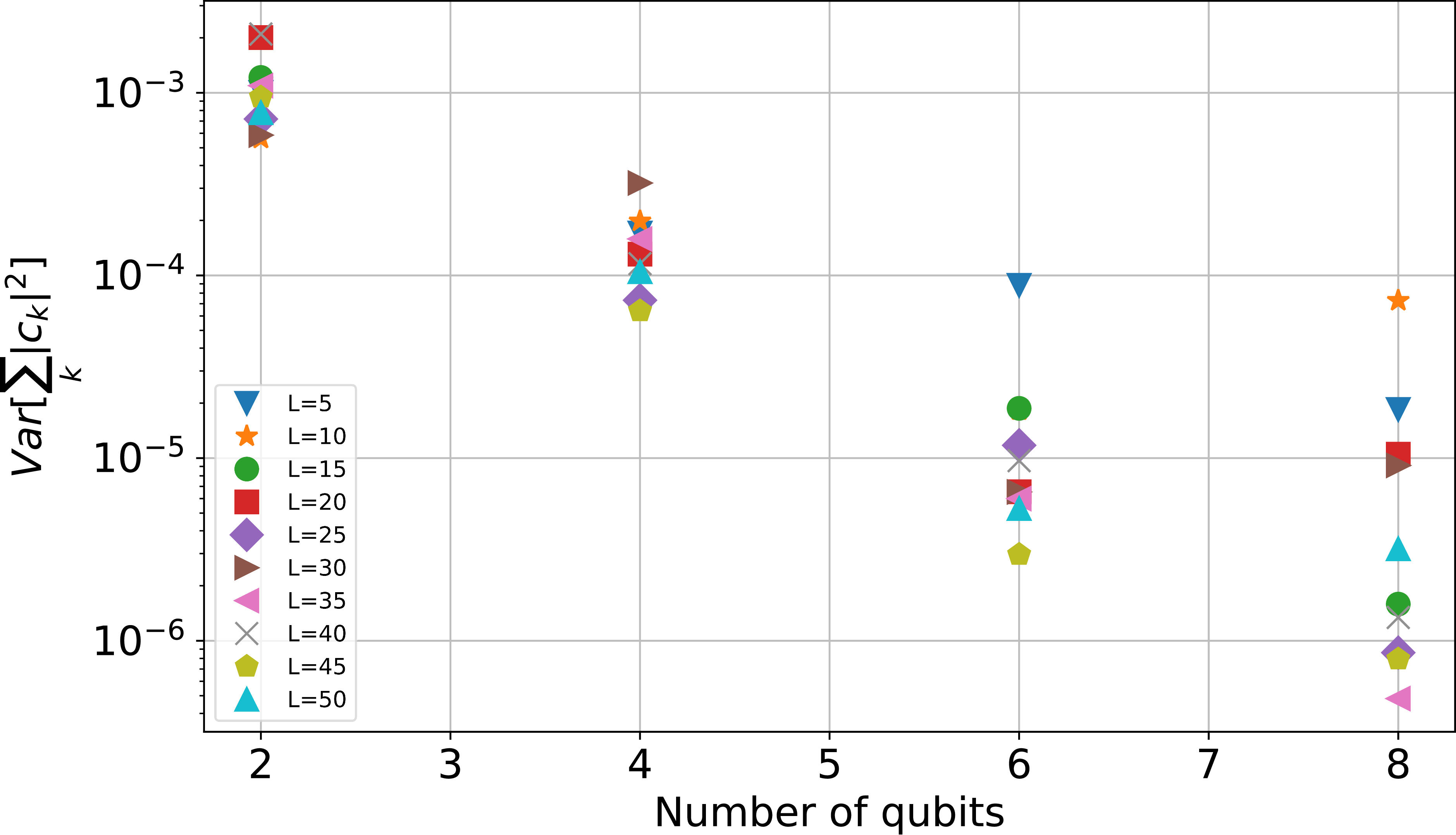}
\end{center}

\caption{(Upper) Mean of squared Fourier coefficients versus $n$ for depths $L\in\{5,10,15,20,25,30,35,40,45,50\}$. The horizontal line shows Eq.~(\ref{fm}). (Bottom) Variance across 300 random seeds.}
\label{res1_2}
\end{figure}

\begin{figure}
\begin{center}
\includegraphics[width=70mm]{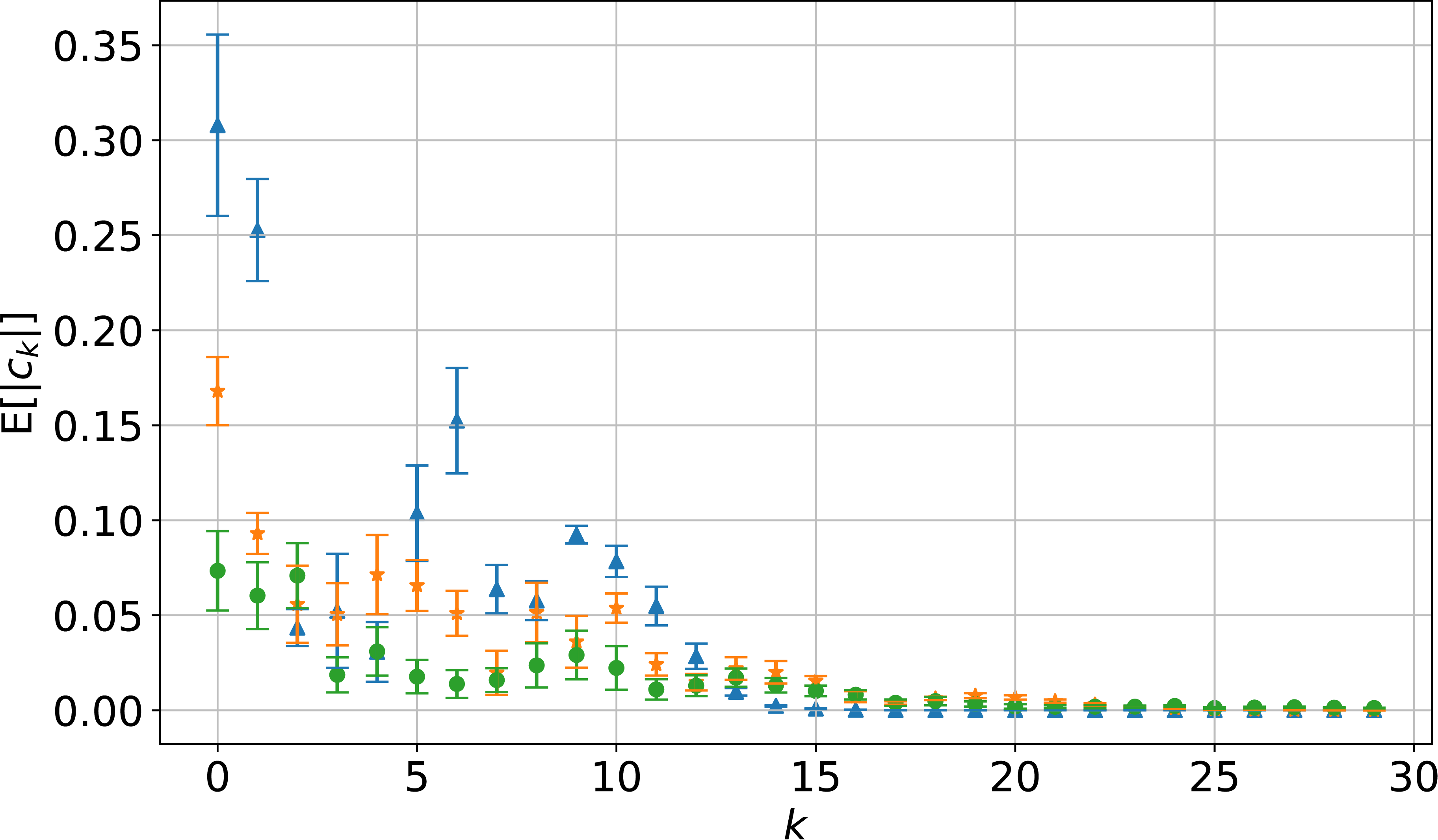}
\end{center}
\caption{Representative Fourier coefficients at $L=15$ and $n\in\{2,4,6\}$. Coefficients decay rapidly with $n$, indicating exponential suppression of spectral power.}
\label{res3}
\end{figure}

Figure~\ref{res1_2} (left) plots the mean spectral power versus the number of qubits $n$ for several depths $L$, with the horizontal line indicating the theoretical 2-design value in Eq.~(\ref{fm}). 
For $L\gtrsim15$, the empirical mean matches Eq.~(\ref{fm}), and the variance across seeds (Fig.~\ref{res1_2}, right) nearly vanishes, indicating concentration. 
This suggests that $E^L(x)$ approaches Haar-like behavior around $L\approx15$ under global measurements. 
To visualize the spectral contraction, Fig.~\ref{res3} shows representative Fourier coefficients at fixed depth $L=15$ for $n\in\{2,4,6\}$; coefficients decay rapidly with $n$, consistent with the exponential suppression predicted by Eq.~(\ref{fm}) and the functional concentration expressed by Eq.~(\ref{f0}). 
Consequently, gradients computed via the parameter-shift rule vanish, yielding a barren-plateau-like landscape from a spectral viewpoint.

Global measurements are known to exacerbate barren plateaus; thus, these observations are conservative regarding trainability. 
When $E^L(x)$ does not approximate a 2-design, Eq.~(\ref{fm}) need not hold, and spectral power deviates accordingly. 
In additional checks (not shown), reducing measurement locality weakened concentration, which is consistent with previous studies.

\begin{figure}[ht]
\[
\underbrace{
\begin{array}{c}
\Qcircuit @C=1em @R=.8em {
   & \gate{R_X(x)} & \gate{R_Y(x)} & \ctrl{1} & \qw      & \qw      & \qw \\
   & \gate{R_X(x)} & \gate{R_Y(x)} & \targ    & \ctrl{1} & \qw      & \qw \\ 
   & \gate{R_X(x)} & \gate{R_Y(x)} & \qw      & \targ    & \ctrl{1} & \qw \\
   & \gate{R_X(x)} & \gate{R_Y(x)} & \qw      & \qw      & \targ    & \qw
}
\end{array}
}_{\displaystyle L \ \text{times}}
\]
\caption{Hardware-efficient embedding $E^L(x)$ for four qubits. Larger $L$ increases encoder expressivity.}
\label{c1}
\end{figure}

\begin{figure}[ht]
\[
\begin{array}{c}
\Qcircuit@C=1em @R=.7em{
\lstick{\ket{0}} &  \multigate{3}{E^L(x)} & \qw   & \gate{R_X(\theta_1)} & \qw & \meter & {Z} \\
\lstick{\ket{0}} &  \ghost{E^L(x)}        & \qw   & \gate{R_X(\theta_2)} & \qw & \meter & {Z} \\
\lstick{\ket{0}} &  \ghost{E^L(x)}        & \qw   & \gate{R_X(\theta_3)} & \qw & \meter & {Z} \\
\lstick{\ket{0}} &  \ghost{E^L(x)}        & \qw   & \gate{R_X(\theta_4)} & \qw & \meter & {Z} \\
}
\end{array}
\]
\caption{QNN used in the experiments (four qubits). The expectation is $f(x,\bm{\theta})=\Tr[U(\bm{\theta})^\dagger E^L(x)^\dagger \rho\,E^L(x)U(\bm{\theta}) Z^{\otimes 4}]$.}
\label{c2}
\end{figure}

\textit{Conclusion.}
We established a direct link between the Fourier representation of parameterized quantum circuits and barren plateaus (BPs): the sum of squared Fourier coefficients—i.e., the total spectral power—decays exponentially with qubit number under BP conditions, providing a spectral explanation for vanishing gradients in variational algorithms. 
Because Parseval’s identity is the second moment under a uniform measure, the analysis is independent of initialization details and requires only parameter periodicity, making it broadly applicable.

Care is needed for very shallow or weakly entangling circuits: insufficient expressivity can render the relevant Fourier expansion uninformative. 
Clarifying this low-expressivity regime may guide BP mitigation via circuit design.

Beyond initialization, BPs may also arise from entanglement growth \cite{30}, noise \cite{31}, and global measurements \cite{23}. 
Understanding how these mechanisms shape the spectral structure identified here is an open direction, with the goal of informing more robust quantum learning architectures.

{\it Acknowledgement.}
This work was supported by JSPS KAKENHI Grant No. 23H01432.
We also received financial support from the programs BRIDGE and SIP (No.~23836436) from Cabinet Office of Japan.

\end{document}